# Decentralized Markets versus Central Control:
# A Comparative Study

**Fredrik Ygge**                                         YGGE@ENERSEARCH.SE
*EnerSearch AB and Uppsala University*
*Chalmers Science Park*
*S-412 88 Gothenburg, Sweden*
*www.enersearch.se/ygge*

**Hans Akkermans**                                 HansAkkermans@CS.VU.NL
*AKMC and Free University Amsterdam*
*Department of Information Management and Software Engineering*
*Computer Science Division*
*De Boelelaan 1081a, NL-1081 HV Amsterdam, The Netherlands*

## Abstract

Multi-Agent Systems (MAS) promise to offer solutions to problems where established, older paradigms fall short. In order to validate such claims that are repeatedly made in software publications, empirical in-depth studies of advantages and weaknesses of multi-agent solutions versus conventional ones in practical applications are needed. Climate control in large buildings is one application area where multi-agent systems, and market-oriented programming in particular, have been reported to be very successful, although central control solutions are still the standard practice. We have therefore constructed and implemented a variety of market designs for this problem, as well as different standard control engineering solutions. This article gives a detailed analysis and comparison, so as to learn about differences between standard versus agent approaches, and yielding new insights about benefits and limitations of computational markets. An important outcome is that "local information plus market communication produces global control".

## 1. Introduction

When new paradigms arise on the scientific horizon, they must prove their value in comparison and competition with existing, more established ones. The multi-agent systems (MAS) paradigm is no exception. In a recent book on software agents (Bradshaw, 1997), Norman observes that perhaps "the most relevant predecessors to today's intelligent agents are servomechanisms and other control devices". And indeed, a number of applications for which multi-agent systems have recently claimed success, are close to the realm of what is traditionally called control engineering. One clear example is the climate control of large buildings with many office rooms. Here, Huberman & Clearwater (1994, 1995) have constructed and tested a working MAS solution based on a market approach, that they reported to outperform existing conventional control.

The key question studied in this article is: *in what respect and to what extent are multi-agent solutions better than their (conventional) alternatives?* We believe that the above-mentioned application provides a nice opportunity to study this question in a detailed empirical way. It is practically very relevant, it lends itself to alternative solutions, and it is quite prototypical for a wide range of industrial applications in distributed resource allocation (including energy management applications (Ygge & Akkermans, 1996; Akkermans & Ygge, 1997; Ygge & Akkermans, 1998), telecoms applications, the file allocation problem of Kurose and Simha (1989), and the flow problems investigated by Wellman (1993)).

This article gives a detailed analysis of a published MAS solution to building climate control, and it compares multi-agent markets with traditional control approaches. We also introduce an improved and novel multi-agent solution to this problem based on an equilibrium market. From our comparative analysis we are able to draw general conclusions about the suitability of the various approaches. Briefly, we show how computational markets can be designed that perform as well as centralized controllers having global information about the total system. However, a major advantage of the market framework is that it achieves this in a fully decentralized fashion using only locally available data. We finally outline that it should be possible to come to a more general theory concerning the connections between markets and conventional control concepts. Here, we show that for the considered type of applications the quasi-equation *"local data + market communication = global control"* holds.

The structure of the article is as follows. After an introduction (Section 2) to market-oriented programming and available theory, Section 3 gives the problem definition. Section 4 introduces the application domain: it describes the office environment and gives the physical model for cooling power and the various temperature and outside weather influences. We then discuss the results of a standard control engineering solution, based on local and independent integral controllers regulating the building climate (Section 5). Next, we review the market-based approach as put forward by Huberman & Clearwater (1994, 1995) (Section 6), and validate their claim that this market approach performs better than conventional independent controllers. We subsequently analyze this market protocol in detail and show that its success is related to the fact that the agents possess global information before the auction is started (Section 7). In Section 8 we then develop an improved standard control engineering scheme that also exploits global data. Such a control scheme turns out to perform much better than the Huberman-Clearwater market. Finally, we propose a market design of our own based on general equilibrium theory. It performs as well as the controller having access to global data, but operates on local data only and thus represents a really decentralized solution (Section 9). Section 10 puts our results into perspective and summarizes the general conclusions in comparing different approaches.

## 2. Market-oriented Programming

The use of market mechanisms for resource allocation by computer systems has a rather long history in computer science, e.g. (Sutherland, 1968), and more recently markets have been used in a number of different application areas.

There is rather extensive theory available for the relations between optimization and markets. For example:

- Jennergren (1973) has shown how a price schedule decomposition algorithm can be used for solving linear programming problems.

- Kurose & Simha (1989) showed how an equilibrium constitutes an optimal solution to a file allocation problem. A similar result was shown by Bertsekas (1992) for the assignment problem.

- Bikhchandani & Mamer (1997) showed that a Pareto efficient outcome[1] maximizes the sum of the utilities in an exchange market with agents holding quasi-linear utility functions[2] This is a significant extension to the above, as it also applies to non-linear problems.[3]

- The results of Bikhchandani and Mamer were extended to markets that include production and uncertainty by Ygge (1998, Chapter 3). The relations between these types of markets and traditional optimization problems were also made more explicit in the latter work.

Markets approaches have been applied to a number of different computerized applications, and some of them are briefly discussed here. Already in 1968 Sutherland proposed an auction mechanism for allocation of computational resources on a PDP-1 computer (Sutherland, 1968). Different amounts of money were assigned to different users in accordance with the importance of their projects. It was reported that the users then over time learn how to bid properly for computational resources. This basic approach has been further refined by, for example, Gagliano et al. (Gagliano et al., 1995).

Kurose & Simha (1989) investigate a file allocation problem. In this work the agents report their marginal utility (for having a certain amount of storage) and its derivative to an auctioneer, which reallocates the resource using a resource-oriented Newton-Raphson algorithm (cf. Ygge & Akkermans, 1998) until an equilibrium has been reached in which the marginal utilities of all agents are the same. Although the paper is based on a number of microeconomic abstractions, it does not for example utilize prices and there is no trade-off between different commodities. So, it is debatable whether this approach can be referred to as really market-based.[4]

The assigning problem of allocating $n$ objects to $n$ users has been investigated by Bertsekas (1992). Each user has a valuation of each object, and cannot be assigned more than one object. It is reported that each user can be seen as an economic agent, and it is shown how an auction (which essentially is an English auction) results in an equilibrium. The assumed agent behavior is that an agent bids $v - w$ for its most highly valued object (i.e. the object with the highest difference, $v$, between the valuation and the price). $w$ is the difference between the valuation and the price of the second most preferred object. There is however no motivation given for the assumed agent behavior, and it is not clear why any agent would use this strategy unless it can be externally imposed. Though there is some economical interpretation of the prices, the approach is not very market-like due to the rather unrealistic assumptions of agent behavior.

---

1. An allocation is Pareto efficient or Pareto optimal if there is no alternative allocation that makes any agent better off without making the outcome worse for some other agent (Varian, 1996, p. 15).
2. An example of a quasi-linear utility function is given in Eq. (19).
3. However, one should remember that this theory does not provide any computational advantages. If it is hard to find the allocation that maximizes the sum of the utilities, then it is also hard to find a Pareto optimal one (as they are the same in this case).
4. One should also note that the formulation of the Newton-Raphson scheme of this approach is overly simplistic, and there are much better standard methods available (Press et al., 1994; Ygge & Akkermans, 1998).

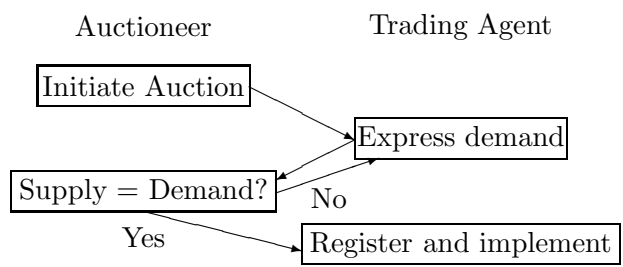

Figure 1: *High-level view of an equilibrium market mechanism.*

The equilibria in the applications by Kurose and Simha, and Bertsekas are proven to be optimal. It is not hard to see that the problem of Kurose and Simha can be reformulated as a proper market with quasi-linear utility functions, and that the competitive equilibrium then is equivalent to the equilibrium described by Kurose and Simha, cf. (Ygge, 1998, Corollary 3.3.1). Similarly, the price vector obtained by Bertsekas clearly constitutes a competitive equilibrium. At the same time, it is shown by Ygge (Ygge, 1998, Theorem 3.2), that all separable optimization problems can be formulated in market terms and the (competitive) general equilibrium (if existing) is identical to the optimal solution to the original optimization problem. Consequently, this recent theory generalizes the earlier theory by for example Kurose and Simha, and Bertsekas.

Wellman et al. have contributed significantly in developing market-based approaches to resource allocation into a programming paradigm, which they have given the name *market-oriented programming*, e.g. (Wellman, 1993; Mullen & Wellman, 1995; Wellman, 1995, 1996; Yamaki, Wellman, & Ishida, 1996; Hu & Wellman, 1996; Cheng & Wellman, 1998; Wellman & Hu, 1998; Walsh, Wellman, Wurman, & MacKie-Mason, 1998). Particularly, the microeconomic framework of general equilibrium theory has been successfully used as a resource allocation mechanism. In such a market which we call an *equilibrium market* agents send *demand functions* telling how much they like to consume or produce at different prices. The auctioneer then tries to establish an equilibrium price vector such that supply meets demand for all commodities, cf. Figure 1.

The process of submitting parts of the demand function may be iterated if an equilibrium price is outside the region captured by the submitted demand functions. One such process is the basic price tâtonnement process, cf. e.g. (Cheng & Wellman, 1998), in which demand functions for the respective commodities are sent to an auctioneer. Each of those demands is based on *expected prices* of the other commodities. That is, if those other prices change, a set of new demand functions may need to be submitted. Once the auctioneer has established an equilibrium price, the agents will exchange the resources as stated by their bid and the equilibrium price. (For example, if an agent states that it wants to buy $1/p$ units of a commodity—where $p$ is the price in the commodity money—and the equilibrium price becomes 1, the agent will buy one unit of resource for one unit of money.)

Equilibrium markets have many attractive theoretical properties. For example, if all agents act competitively[5], the outcome is Pareto efficient. Furthermore, if all utility functions are quasi-linear, the outcome is globally optimal (Ygge, 1998, Theorem 3.2), and in the presence of uncertainty, the outcome maximizes the expected global utility (Ygge, 1998, Theorem 3.5). We note that equilibrium markets be computationally implemented in a computationally very efficient manner (Ygge, 1998, Chapter 4).

Wellman et al. have applied equilibrium markets to a number of applications, such as multi-commodity flow problems, design problems, and bandwidth allocation problems. They have also introduced a market-based approach to scheduling which has many similarities with the assignment problem of Bertsekas described above, but relies on more realistic assumptions on agent behavior.

The present authors have introduced a market-oriented approach to power load management (Ygge & Akkermans, 1996; Ygge, 1998; Ygge et al., 1999).

We note that the aim of market-oriented programming in computer science is fundamentally different from the aim of economic theory. This is visualized in Figure 2. In market-oriented programming, microeconomic theory is taken as given and serves as the theory for implementation of computational agents. Whether or not the microeconomic theory actually reflects human behavior is not the critical issue. The important question is instead how microeconomic theory can be utilized for the implementation of successful resource allocation mechanisms in computer systems. For example, even though no (or at least very few) people believe that humans use explicit utility functions when making their decisions, such functions do appear very useful to concisely represent human preferences for use in computational agents.

It is obviously very interesting to investigate the use of computational markets for automating trades between different self-interested parties, where information is private and is revealed only if there is an expected gain from doing so. However, the use of markets has also been proposed for standard resource allocation where the true utility/costs for all nodes that consume and produce resource is assumed to be available (though possibly uncertain and/or distributed in the system). The main arguments found in the literature for applying market to these types of problems are:

- The numerous similarities between economic systems and distributed computer systems suggest that models and methods previously developed within the field of mathematical economics can serve as *blueprints* for engineering similar mechanisms in distributed computer systems (Kurose & Simha, 1989).

- Auction algorithms are highly intuitive and easy to understand; they can be explained in terms of economic competition concepts, which are couched on everyday experience (Bertsekas, 1992).

---

5. An agent that acts competitively treats prices as exogenous, that is, the impact on the prices due to its own behavior is negligible (Varian, 1996, p. 516). This is a very reasonable assumption if the market is of at least moderate size and/or if there is uncertainty about the behavior of the other agents (Sandholm & Ygge, 1997).

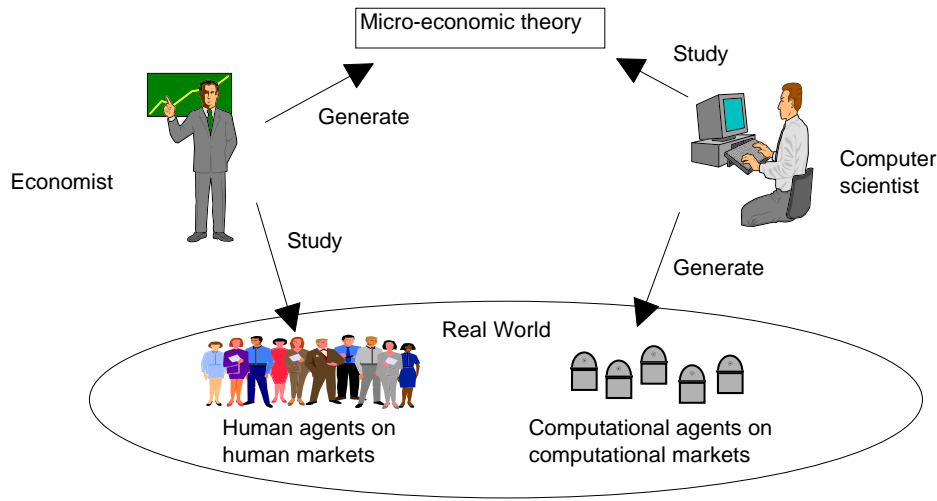

Figure 2: *A simplified view on the relation between economics and computer science with respect to microeconomic theory. Economists study humans and generate theory that is used for the explanation of human economic behavior. Computer scientists doing market-oriented programming use this theory as a basis for building working computational market systems.*

- Market approaches enable a natural decomposition, both from a software engineering perspective as well as from a computational perspective (Schreiber et al., 1999; Ygge, 1998, Chapter 15).

- Market approaches are very flexible in that they allow for ongoing addition and deletion of agents. No global changes are required—merely the demand/supply relation is altered (Ygge, 1998, Chapter 15).

- Markets are informationally efficient in terms of information dimensionality (Jordan, 1982), and the abstractions used are the most natural ones for the user (Ygge, 1998, Chapter 15).

- The introduction of trading resources for some sort of money enables evaluation of local performance and valuation of resources, so that it becomes apparent which resources are the most valuable and which agents are using the most of these (Ygge, 1998, Chapter 15).

The above arguments are mainly conceptual and related to software/system design and engineering issues, and must prove their value from acceptance by software and system designers. But there are also different and in some respects stronger claims. For example, Huberman and Clearwater state (Huberman & Clearwater, 1995) for the application of building control that: *"While in principle an omniscient central controller with access to all the environmental and thermal parameters of a building (i.e. a perfect model) could* *optimally control it, in practice such knowledge is seldom available to the system. Instead, partial information about local changes in the variables (such as instantaneous office occupancy, external temperature, and computer use) is the only reliable source that can be used for controlling the building."* As an alternative to such an omniscient controller Huberman and Clearwater propose a market-based multi-agent approach to this problem. This raises a very interesting question: Are there applications in which the information structure is such that market-based approaches do better than traditional approaches? In this paper we will carefully examine this issue for a building climate control problem. We will investigate what information different alternatives (traditional and new market-based) require, and what the performance of these different approaches are. We will use exactly the same problem formulation as Huberman and Clearwater in order to reproduce their results and evaluate other new approaches with their original formulation. This problem formulation is given in the next section.

## 3. Problem Definition

The task is to allocate a resource (cold air) in an office building, given the setpoint temperatures of the respective offices. As a measure of the success of the allocation, the standard deviation of the deviation from the setpoint is used (Huberman & Clearwater, 1995), i.e.

$$StdDev(T_i - T^{setp}) = \sqrt{\frac{1}{N} \sum_{o=1}^{N} [(T_{io} - T_o^{setp}) - (\langle T_i \rangle - \langle T^{setp} \rangle)]^2}, \qquad (1)$$

where $\langle \cdot \rangle$ denotes the average value of a variable, $T_{io}$ is the actual temperature of an office, and $T_o^{setp}$ is the setpoint temperature. The index $i$ denotes a time interval under observation and the index $o$ denotes the office under observation. This naming convention is used throughout the article. (In addition the index $k$ will also be used to denote time periods.)

It may be debated whether or not this is the best measure, but we stick to it in this article in order to evaluate the approach taken by Huberman and Clearwater using their own measure.

## 4. The Office Environment

The case study we consider for the comparison of decentralized markets versus central control solutions is the building climate control of a large office environment.

In this section, we present a mathematical-physical model of the office environment. We first give a conceptual summary so that it is possible to understand the basic ideas of the model without studying the equations. The offices are attached to a pipe in which the resource (cold air) is transported as in Figure 3. The characteristics of this system are similar to the characteristics of a district heating system, but with offices instead of households. We assume that there are 100 offices in total, and that they are equally distributed towards East, South, West, and North.

The thermodynamics of the office environment is actually quite simple. Every office is seen as a storage place for heat, but heat may dissipate to its environment. In the model, the thermodynamic behavior of an office is equivalent to a basic electrical RC-circuit. Here, voltage is analogous to temperature, and electrical current is analogous to heat flow. $C$ and $R$ then respectively denote heat capacitance and thermal resistance.

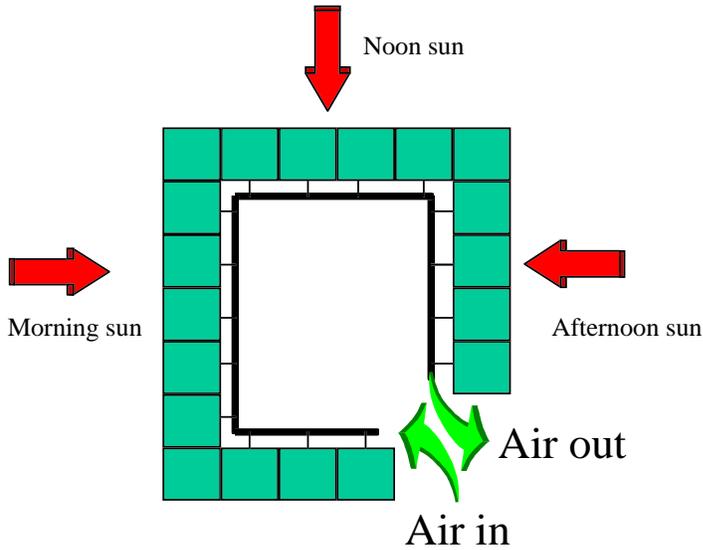

Figure 3: *Offices and air transportation.*

A good general reference on thermodynamic models as the one we use here is the book by Incropera & Witt (1990). The heat equations are continuous in time, but are discretized according to standard procedures from control engineering, cf. (Ogata, 1990). The ontology and reusability aspects involved in thermodynamics model construction are discussed extensively by Borst, Akkermans, & Top (1997).

## 4.1 Thermodynamic Properties

The resource treated is cooling power. Each office can make use only of a fraction, $\eta$, of the available resource at that office, $P_{io}^{avail}$, so that

$$P_{io}^{cons} \leq \eta \cdot P_{io}^{avail}, \tag{2}$$

where $P_{io}^{cons}$ is the consumed power. The available resource at one office is equal to the available resource at the previous office minus the consumed resource at the previous office. Throughout this article we assume an $\eta$ of 0.5.

We treat everything in discrete time. The time interval we use in the calculations is one minute. For all offices the temperature, $T_{io}$, is obtained by integrating a differential equation in discretized form:

$$T_{io} = T_{0,o} + \sum_{k=1}^{i}(P_{ko}^{heat} - P_{ko}^{cons})/C_o, \tag{3}$$

where $P_{ko}^{heat}$ is the heating power and $C_o$ is the thermal capacitance. The heating power is described by

$$P_{io}^{heat} = (T_{io}^{virt} - T_{io})/R_o, \tag{4}$$

where $R_o$ is the thermal resistance and $T_{io}^{virt}$ is a virtual outdoor temperature, described in more detail below.

From Eqs. (3) and (4) we see that there is a feedback loop between the office temperature and the heating power. Solving for the temperature we obtain

$$T_{io} = \frac{1}{1 + \frac{1}{R_o C_o}} \left( T_{i-1,o} + \frac{\frac{T_{io}^{virt}}{R_o} - P_{io}^{cons}}{C_o} \right), \, i > 0. \tag{5}$$

This is the equation for the dynamics of the system that can be directly computed. At the right-hand side we have known quantities, where $C_o$ and $R_o$ are externally given building parameters, $P_{io}^{cons}$ is the output of the utilized controller (as will be described in detail later in the article for various different controllers), and $T_{io}^{virt}$ is obtained from the weather model below.

## 4.2 Weather Model

All external weather influences on the office environment are modeled by a virtual temperature, representing the outdoor temperature, sun radiation, etc. We assume that there is sunshine every day and that the outdoor temperature, $T^{outd}$, varies from 22 to $35°C$ according to

$$T_i^{outd} = 22 + 13 \cdot e^{-((i \cdot s - 4) \, mod \, 24 - 12)^2 / 20}, \tag{6}$$

where $s$ is the length of each time interval expressed in hours, so here $s = 1/60$.

The virtual temperature, $T_{io}^{virt}$, is described by

$$T_{io}^{virt} = T_i^{outd} + T_o^{sun} + T_{io}^{fluct}, \tag{7}$$

where $T^{fluct}$ is a random disturbance, thought to represent small fluctuations caused by for example the wind. $T^{fluct}$ is Gaussian distributed with zero mean and a standard deviation equal to unity. $T^{sun}$ is the sun radiation component. For the offices located at the East side $T^{sun}$ is described by

$$T_i^{sun,East} = 8 \cdot e^{-((i \cdot s + 4) \, mod \, 24 - 12)^2 / 5}, \tag{8}$$

and correspondingly for the South and the West offices

$$T_i^{sun,South} = 15 \cdot e^{-(i \cdot s \, mod \, 24 - 12)^2 / 5}, \tag{9}$$

and

$$T_i^{sun,West} = 8 \cdot e^{-((i \cdot s - 4) \, mod \, 24 - 12)^2 / 5}. \tag{10}$$

The various temperatures are plotted in Figure 4.

## 4.3 Office Temperatures without Control

In Figure 5, the temperature for a South-oriented office is plotted with different thermal resistances, $R_o$, and thermal capacitances, $C_o$. For simplicity we assume all $R_o$ to be equal and all $C_o$ to be equal. From this figure we observe two things: first, the higher $R_o C_o$ the bigger the lag in temperature changes, and second, the higher $R_o C_o$ the smaller the fluctuations in temperature. For the simulation experiments in this article we took $R_o = 10$ and $C_o = 10$. An $R_o$ or a $C_o$ equal to zero implies that $T_{io} = T_{io}^{virt}$, as can be seen by letting $R_o$ or $C_o$ approach zero in Eq. (5). Clearly, without control the office temperatures strongly fluctuate and reach unacceptably high values in all cases.

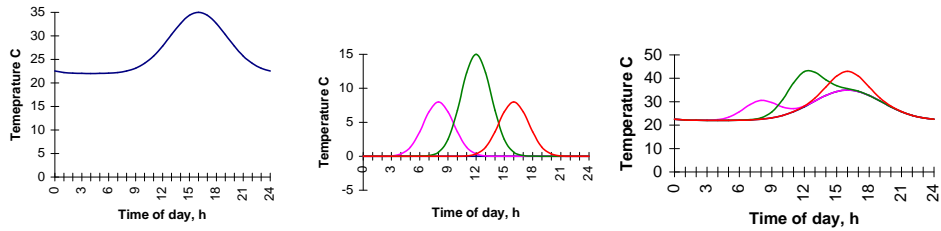

Figure 4: *The plot at the left shows the outdoor temperature, $T_i^{outd}$. The middle plot shows the sun radiation components, $T^{sun}$, (with peaks at time 8, 12, and 16 hours for offices located at the East, South, and West sides, respectively). Finally, the outdoor temperature plus the sun radiation components are plotted at the right.*

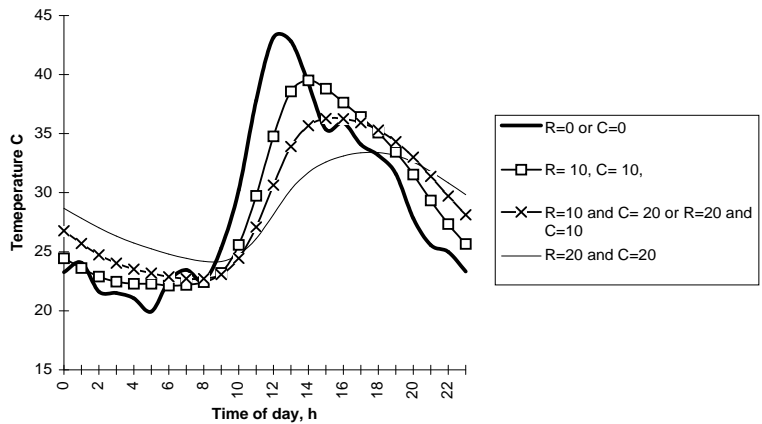

Figure 5: *The indoor temperature for an uncontrolled office is plotted for different values of the thermal resistance and the heat capacitance. Small values of the thermal resistance and capacitance give higher peaks, while higher values give smoother curves.*

## 5. Control-A: **Conventional Independent Controllers**

### 5.1 Integral Control for the Offices

The application of regulating office temperature has a long tradition in control theory. The most widely known controllers are different variants of the PID controller. The letters PID denote that the control signal is proportional (P) to the error (that is, to the difference between the setpoint and the actual value); proportional to the integral (I) of the error; or proportional to the derivative (D) of the error. Here, we use a variant of an integrating controller[6] of the form

$$F_{io} = F_{i-1,o} + \beta(T_{io} - T_o^{setp}),$$  (11)

where $F$ is the output signal from the controller, and $\beta$ is a so-called gain parameter (that can be set externally in the design of the controller). For the simulations it is assumed that $F_{io}$ is limited to a value between zero and three. This is in order to model that the valves can be closed but that cooling resources are not delivered from one room to another (the lower bound), and that there is a maximum to the amount of resources that can be obtained by opening the valve fully (the upper bound). The control signal $F_{io}$ is sent to the actuator and the actual $P_{io}^{cons}$ is obtained from

$$P_{io}^{cons} = \begin{cases} F_{io}, & F_{io} \leq \eta \cdot P_{io}^{avail} \\ \eta \cdot P_{io}^{avail}, & F_{io} > \eta \cdot P_{io}^{avail} \end{cases}.$$  (12)

Plots of the office temperatures with different gains are shown in Figure 6. The gain of the controller is not critical for this application. Too high a gain will result in the controller overreacting when the temperature exceeds the setpoint, after which it will be under the setpoint for quite some time. This leads to a larger error than if smaller adjustments are made. Also, the amplitude of the control signal then gets unnecessarily high, but the system does not get dangerously unstable. We note that the maximum deviation here is $\pm 0.06°C$. Thus, controllers using any of the three gains perform very well. In the further calculations of this article a gain equal to 10 has been adopted.

### 5.2 The Implications of Limited Resources

So far, we have assumed that the total amount of available resources is unlimited. Now, we suppose that there is a maximum value for the cooling power that is inserted into the system. In such a situation, offices that are situated close to the air input will obtain a sufficient amount of cool air, but those near the end will suffer if totally uncoordinated controllers are used. Thus, the smaller the total amount of available resources, the larger the standard deviation will be. This is visualized in Figure 7. As a reasonable figure we have chosen an upper limit for the total resource amount of 140.

We conclude, as shown by the example, that independent integrating controllers perform very well when the amount of cooling resources is unlimited. On the other hand, when there is a shortage of available resources, the standard deviation increases dramatically, yielding poor performance.

---

6. The main reason for choosing an integrating controller is that we want to have a setting that is as close as possible to the setting described by Huberman and Clearwater (1995). Other controllers for this application may well be considered, but as long as the performance is as good as shown in Figure 6, this is not at all crucial for the central argument of this article.

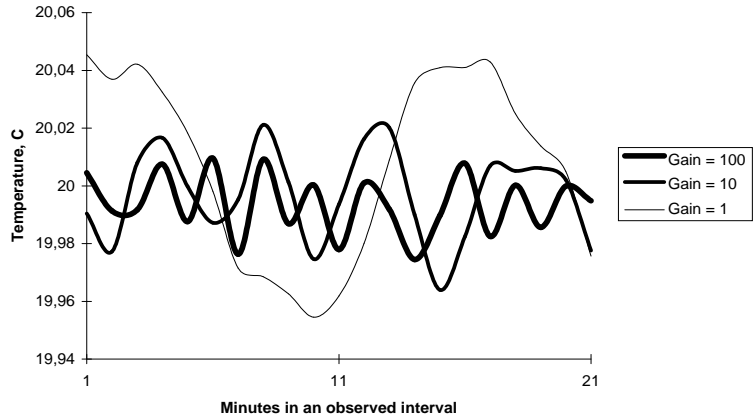

Figure 6: *The indoor temperature for an office, utilizing an integral controller, is plotted for different controller gains. The setpoint temperature is 20°C.*

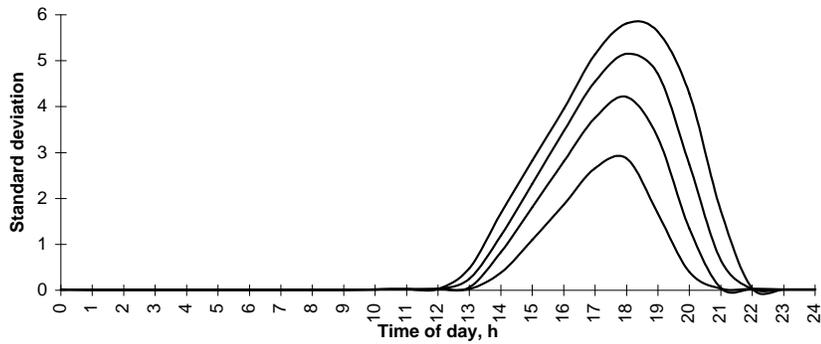

Figure 7: *The standard deviation (in °C) , as defined in Eq. (1) for each interval, i, is plotted for different amounts of available cold air resources (130, 140, 150, and 160) with 100 offices. The lower the amount of available resources, the higher the standard deviation.*

## 6. Market-A: The Approach by Huberman and Clearwater

A multi-agent systems solution to the problem of building control has been presented by Huberman and Clearwater (1994, 1995). The approach taken is to model the resource allocation problem of the building environment as a computational market where agents buy and sell cooling power resources. The non-separability in terms of agents is ignored.[7]

The basic idea is that every office is represented by an agent that is responsible for making good use of its resources, and to this end trades resources with other agents. The agents send bids to an auctioneer, which calculates a clearing price, and makes sure that no agent has to buy for a price higher than its bid nor has to sell for a price lower than its bid.

In this section we give the market protocol proposed by Huberman and Clearwater, and we reproduce their results. In order to make the paper self-contained, the interested reader can find the original equations underlying the Huberman-Clearwater market protocol in Appendix A.

### 6.1 Market protocol

A bid is constructed as the following tuple:

$$bid = [sell_{io}, v_{io}, B_{io}], \tag{13}$$

where $sell$ is a boolean variable indicating whether the current bid is a $sell$ bid ($true$) or buy bid ($false$), $v$ is the volume traded for, and $B$ is the price that the agent demands (sell bid) or is prepared to pay (buy bid).

Each of the variables in Eq. (13) is a function of other variables according to

$$\begin{aligned} sell_{io} &= sell_{io}(T_{io}, T_o^{setp}, \langle T_i \rangle, \langle T_o^{setp} \rangle), \\ v_{io} &= v_{io}(\mathbf{T}_i, \mathbf{T}^{setp}, \alpha), \text{and} \\ B_{io} &= B_{io}(T_{io}, T_o^{setp}, \langle T_i \rangle, \langle T_o^{setp} \rangle, m_{io}), \end{aligned} \tag{14}$$

where $\mathbf{T}_i = [T_{i1}, T_{i2}, \ldots, T_{i100}]$, $\mathbf{T}^{setp} = [T_1^{setp}, T_2^{setp}, \ldots, T_{100}^{setp}]$, $\alpha$ is a strength parameter of the auction, and $m_{io}$ is called a money parameter. The trade volume is directly proportional to $\alpha$ and, hence, doubling $\alpha$ doubles the traded volume in an auction. Thus, $\alpha$ is a free parameter that can be externally set to tune the trade volumes to proper levels. The money parameter does not correspond to any real currency; it is merely a control variable, varying between 100 and 200, and it does not have any direct connection to, for example, the value of the cooling resource. The money cannot be saved between auction rounds in the Huberman-Clearwater protocol, another indication that it simply functions as an adjustment parameter in this protocol.

---

7. A resource allocation problem is separable if the total utility of the system of the system can be expressed as the sum of the utilities of each node (producer/consumer), and the only constraints on how much resource that can be assigned to each node is only restricted by that their sum must not exceed the total available resource (Ibaraki & Katoh, 1988). Hence, the problem treated in this paper is not separable as there are other constraints on how the resource can be distributed. For example, if we assign sufficiently much to the penultimate agent, the ultimate will have a significant amount of resource available.

An auction consists of the following three phases:

1. All agents send their bids, $[sell_{io}(T_{io}, T_o^{setp}, \langle T_i \rangle, \langle T_o^{setp} \rangle,$ $v_{io}(\mathbf{T}_i, \mathbf{T}^{setp}, \alpha),$ $B_{io}(T_{io}, T_o^{setp}, \langle T_i \rangle, \langle T_o^{setp} \rangle, m_{io})]$, to the auctioneer.

2. The auctioneer then computes a market clearing price by searching for the following minimum:

$$\min_{p_i} \left| \sum_{o|sell=true, B_{io} \leq p_i} v_{io} - \sum_{o|sell=false, B_{io} \geq p_i} v_{io} \right|. \tag{15}$$

   This yields a price such that supply matches demand as closely as possible.

3. Finally, the resources are allocated as dictated by the above bids and the obtained market price, according to the rule: all sellers that are offering $v_{io}$ for a price, $B_{io}$, that is lower than or equal to the market price will sell $v_{io}$; and correspondingly for buyers.[8]

As an example of the computation of the equilibrium price, assume that the auctioneer has received the following bids $[true, 2, 4]$, $[true, 1, 3]$, $[true, 2, 2]$, $[false, 1, 3]$, $[false, 2, 2]$, $[false, 2, 1]$. Then the market clearing price is 3, leading to acceptance of bid $2, 3, 4$, and 5. (Bid 1 is too high a sell price and bid 6 is too low a buy price.)

## 6.2 Simulations

Figure 8 shows two plots of a simulation for the period between 3 p.m. and 7 p.m.[9] The initial temperatures for all offices were set to $20°C$. The upper plot is the standard deviation when independent integrating controllers are used, and the lower one shows the agent-based control scheme as defined above. We found that for the adjustment parameter $\alpha$ (see Eq. (23)) a value of 64 led to the smallest overall standard deviation. The key observation from the figure is that the agent approach offers at least one order of magnitude of improvement.

Compared to independent conventional controllers this is indeed a major advance, and this is the central claim made by Huberman and Clearwater. The market approach given above leads to a reduced control error and thus to increased comfort compared to standard controllers. Our simulations thus reproduce and validate the results of Huberman & Clearwater (1994, 1995).

---

8. Since the bids are given using discrete volumes, supply will seldomly match demand *exactly* at the clearing price. Normally, there will be a very small excess demand or supply. If there is an excess supply, all buyers that are willing to pay at most the clearing price will buy, but not all sellers that are willing to accept at least the clearing price can sell. In this situation, a seller is selected randomly from the valid candidates does deliver only a fraction of its bid. A corresponding procedure is used when there is a small excess demand.

9. All solutions discussed in this article have been implemented in and simulated by the authors using C++ on a PC running Windows95. Furthermore, all simulations have been independently recreated and verified by Bengt Johannesson from various relevant papers (Ygge & Akkermans, 1997; Huberman & Clearwater, 1995; Clearwater & Huberman, 1994)), as part of a master's thesis project, using Python on a PC running Linux.

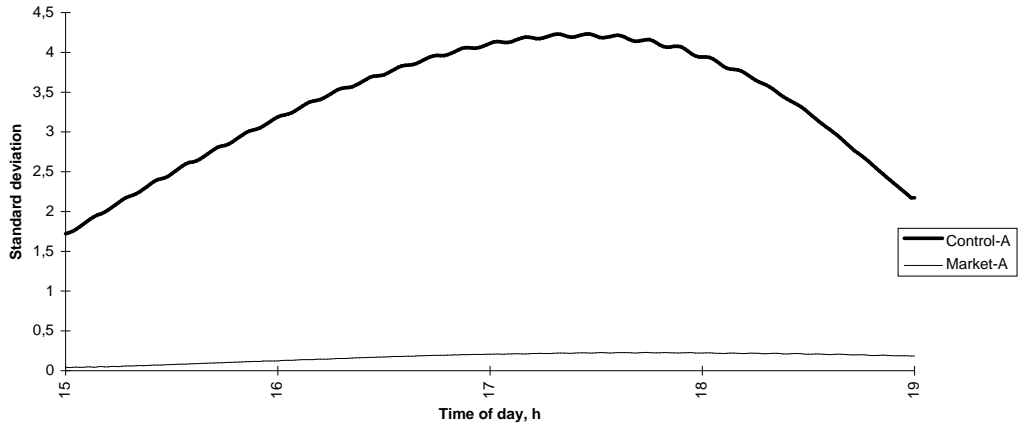

Figure 8: *Standard deviation for independent controllers (top) and agent-based control (bottom).*

However, this cannot be the last word in a successful system study. A first question to be raised is whether we can find out in more detail what the actual reason is for the success of the above market-based approach. A second relevant question is whether there are alternative, market-based and/or central control-based, solutions to the building control problem that are even better. The answer to both questions is *yes*, as we set out to demonstrate now.

## 7. Market-A′: A Suite of Variations

In this section we present a suite of variations on the scheme presented in Section 6. The main aim is to understand which of many factors involved are actually responsible for the good performance of the Huberman-Clearwater market approach.

### 7.1 Deleting the Money Dependency

A first simplification is to remove the adjustment parameter called money. This is done by setting $m_{io}$ to a very high value, $C$, regardless of the resource allocation.[10] The simplified market protocol then is:

1. All agents send their bids, $[sell_{io}(T_{io}, T_o^{setp}, \langle T_i \rangle, \langle T_o^{setp} \rangle, \quad v_{io}(\mathbf{T}_i, \mathbf{T}^{setp}, \alpha),$
   $B_{io}(T_{io}, T_o^{setp}, \langle T_i \rangle, \langle T_o^{setp} \rangle, m_{io} = C)]$, to the auctioneer.

2. The auctioneer computes a market price from

$$\min_{p_i} \left| \sum_{o \mid sell = true, B_{io} \leq p_i} v_{io} - \sum_{o \mid sell = false, B_{io} \geq p_i} v_{io} \right|. \qquad (16)$$

3. Allocate the resource as dictated by the bids and the market clearing price.

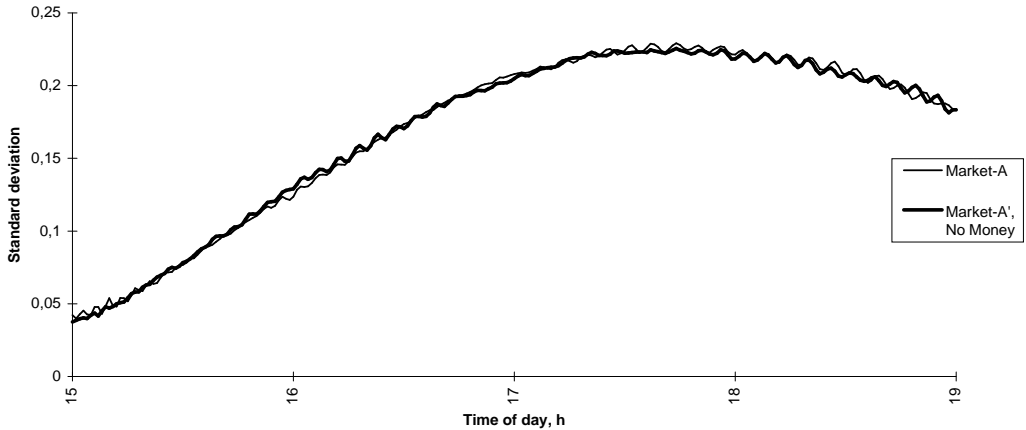

Figure 9: *Standard deviation for the original Huberman-Clearwater scheme and a scheme without any money dependencies.*

Plots of the standard deviation with the original scheme as well as with a scheme where all money dependencies have been removed are shown in Figure 9. For the scheme without money, setting the $\alpha$ parameter to 66 turned out to be optimal. The scheme without money dependencies performs as well as the original scheme. The reason for this is that $B_{io}$'s dependency on $m_{io}$ is completely negligible.[11] Hence, the money parameter does not play any significant role in the success of the Huberman-Clearwater market scheme.

## 7.2 Deleting the Temperature Dependency

Another factor of interest is the impact of the temperature on the bid prices. We remove this dependency by setting the bid price to 10 for all selling agents and to 100 for all buying agents. That is, we now use the following market scheme:

1. All agents send their bids,

   $[sell_{io}(T_{io}, T_o^{setp}, \langle T_i \rangle, \langle T_o^{setp} \rangle), v_{io}(\mathbf{T}_i, \mathbf{T}^{setp}, \alpha), B_{io} = \left\{ \begin{array}{ll} 10, & sell = true \\ 100, & sell = false \end{array} \right. ]$,

   to the auctioneer.

2. Now, all prices larger than 10 and smaller than 100 are equally good candidates. If there is a mismatch between supply and demand, say, supply exceeds demand, the agents that will sell are picked randomly among the valid candidates, and the resources are allocated accordingly.

---

10. More precisely, this is done by setting $U(0, m)$ in Eq. (28) (Appendix A) to a constant value, 2000. This corresponds to letting $m_{io}$ approach infinity.

11. In actual fact, $U(0, m)$ in Eq. (28) of Appendix A will vary only between 1999.86 and 2000 for all possible $m_{io}$.

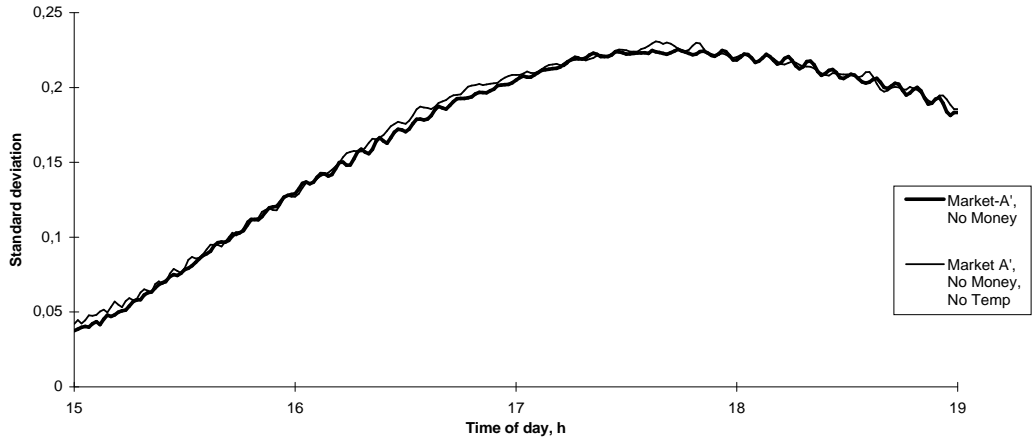

Figure 10: *Standard deviation for the scheme with the money dependencies removed, compared to a scheme with both the money and temperature dependencies removed.*

In Figure 10 the standard deviation is plotted for the scheme without money, mentioned above, and for a scheme where the dependency on the temperature has been removed as well. Here, setting the fit parameter $\alpha$ to 65 turned out to be optimal. We see that also here the performance is as good as that of the original scheme. Note that the temperature is still used to determine both $v_{io}$ and the boolean variable *sell*. But the conclusion is that the temperature dependency does not affect the quality of the market-based solution to the building control problem.

### 7.3 Deleting the Auction

Next, we let the agents assign their bids to themselves *without any auction whatsoever*. This means that the sum total of the controller outputs, $F_{io}$, might sometimes exceed the total resource amount and sometimes be below that. The physical model is of course still obeyed, so that the total resource amount actually used, the total consumed cooling power $P_{io}^{cons}$, will never exceed the totally available resource amount, as described by Eq. (2). Hence, the resource updates are described by:

$$F_{io}' = F_{i-1,o} \begin{cases} +v_{io}(\mathbf{T}_i, \mathbf{T}^{setp}, \alpha), & sell = false \\ -v_{io}(\mathbf{T}_i, \mathbf{T}^{setp}, \alpha), & sell = true \end{cases}$$
$$F_{io} = \begin{cases} 0, & F_{io}' < 0 \\ F_{io}', & 0 \leq F_{io}' \leq 3 \\ 3, & F_{io}' > 3 \end{cases} \tag{17}$$

The result of this simulation is shown in Figure 11. Here, an $\alpha$ of 17 turned out to be optimal. The surprising conclusion is that the multi-agent scheme *without* performing any auction is roughly ten times better than the auction-based MARKET-A scheme.[12]

---

12. From a practical point of view this is a pathological solution, revealing a loophole in the problem definition of Clearwater and Huberman, cf. Section 3. The revealed loophole of the problem definition is

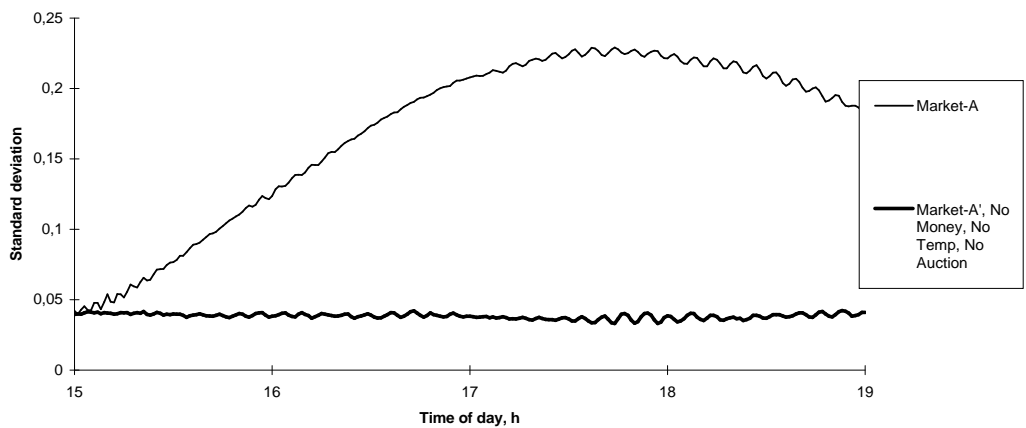

Figure 11: *Standard deviation with the auction mechanisms completely removed, compared to the original Huberman-Clearwater* MARKET-A *scheme.*

## 7.4 Discussion

At first glance, it might seem counterintuitive that performance actually *improves* significantly when the core mechanisms of a market are removed. First we showed that introducing the market improves performance considerably compared to conventional independent control, and then we showed that market mechanisms are superfluous. What, then, is the big difference between the uncoordinated integrating controller, the CONTROL-A scheme, and the multi-agent scheme without any auction that we ended up with? The simple answer, in our view, is the *access to global data that each agent has*, in terms of the *average* temperature and the *average* setpoint temperature (as follows from Eq. (14)), in the MARKET-A and MARKET-A′ schemes. Knowing average quantities means that an agent has some information about the *other* agents c.q. offices. This information is not available in the case of independent conventional controllers. The exploitation of this non-local information makes the big difference.

---

that it does not take absolute temperature into account, but only the differences between temperature (i.e. minimizing standard deviation while potentially sacrificing average temperature). In the proper approaches introduced in the remainder of this article we do not allow ourselves to take advantage of this loophole, but minimize differences while using all available resource. That is, all approaches that are introduced from now on solve a problem which is strictly harder than the problem defined in Section 3. What happens in the Market-A scheme without an auction can be seen from observing the definition of $t_{io}$ in Eq. (22). It is based on a quotient between the setpoint temperature and the actual temperature. Then the request volume, Eq. (24), is proportional to this quotient. This always results in a significant excess supply. (Hint, think of a small example with two offices, both with setpoint 20, and with temperatures 19 and 21. Now, $\frac{20}{19}$ is farther away from 1 than $\frac{20}{21}$, and hence the selling volume will exceed the buying volume.) This implies that at each trade occasion, many offices are prevented from selling (because of poor definitions of the volumes). The offices that are prevented to sell will deviate significantly, and hence this results in a high value of the measure. When the auction is deleted, all agents that are below the average distance to the setpoints are now able to get rid of the resource and hence the differences in temperature will decrease.

Thus, the upshot of our factor analysis of the success of the Huberman-Clearwater market scheme of Section 6 is that the agents have to share crucial office and control information first, before the auction takes place. This is why one can remove the auction without any deterioration of the results: a major function of an auction is to share necessary information through the bidding procedure, but in the Huberman-Clearwater protocol the needed information has already been shared. A more appropriate description of this market scheme would therefore be to say that it consists of not three (as described in Section 6), but four phases, in which the first phase is that all agents send their current temperatures $T_{io}$ as well as their setpoint temperatures $T_o^{setp}$ to all other agents.

In sum, we have now answered the first question raised at the end of the previous section: What is the reason for the success of the Huberman-Clearwater market scheme, compared to conventional independent controllers? The single most important factor is contained in the first step above: *before* the market procedure is carried out, the agents communicate key information about their current situation to all other agents. If that is done neither money nor auction mechanisms are needed. For conventional independent controllers, with which Huberman and Clearwater compare their own approach, this globally shared information is *not* available, however. This explains why these two solutions perform so very differently.

Now, we turn to the second question raised in the previous section: are there alternative and perhaps better solutions? The above analysis suggests that we look into two different directions. If one wants to persuade control engineers of the value of agent-based approaches, it will be more convincing to compare the Huberman-Clearwater scheme not to strictly local controllers, but to controllers that have access to the same global information as the agents have. Secondly, this dependency on global information by the agents is a highly undesirable feature that clearly runs counter to a decentralized computational market philosophy. So, one should investigate whether it is possible to design a market protocol that exploits only strictly local information in the agents' bid construction. These issues we investigate in the following two sections, first treating central control and subsequently devising a strictly local market protocol. It will be seen that both yield significantly improved performance.

## 8. Control-B: A Standard Controller Using Global Data

Having concluded that the access to global data is crucial for performance, it is of course of interest to analyze what the performance would be of an integrating controller, like the one introduced in Eq. (11), *but now incorporating global data.*

We would like the controller to take into account not only its own deviation from its setpoint, but to consider also the deviations of the other offices from their setpoints. Therefore, the controller in Eq. (11) is extended to

$$F_{io} = F_{i-1,o} + \beta[(T_{io} - T_o^{setp}) - (\langle T_i \rangle - \langle T^{setp} \rangle)], \qquad (18)$$

where $\beta$ is set to 10 as previously. $P_{io}^{cons}$ is, as before, obtained from Eq. (12).

The plot from the simulation with this controller, compared to the Market-A simulation, and to the Market-A$'$ simulation where the auction was removed, is shown in Figure 12.

We see that the standard deviation is approximately the same for the Control-B and the Market-A$'$ schemes. Thus, the Control-B scheme also performs roughly ten times better than Market-A. An important difference, though, is that Control-B employs a well-known integrating controller for which there are well-understood methods and theories for e.g. stability analysis.[13] In contrast, the Market-A scheme is not easily analyzable from a formal theory perspective, since it does not rely on such well established concepts. While Control-B is relatively clear, see Eq. (18), it is much harder to conceptually grasp the principles behind Market-A, as it consists of a large number of complicated and not easily justified equations (see Eqs. (21)–(33) in Appendix A). The main conclusion from the present section is that, if we enable conventional controllers to act upon the same global information as the Huberman-Clearwater agents do, the central control approach performs best and is easiest to understand.

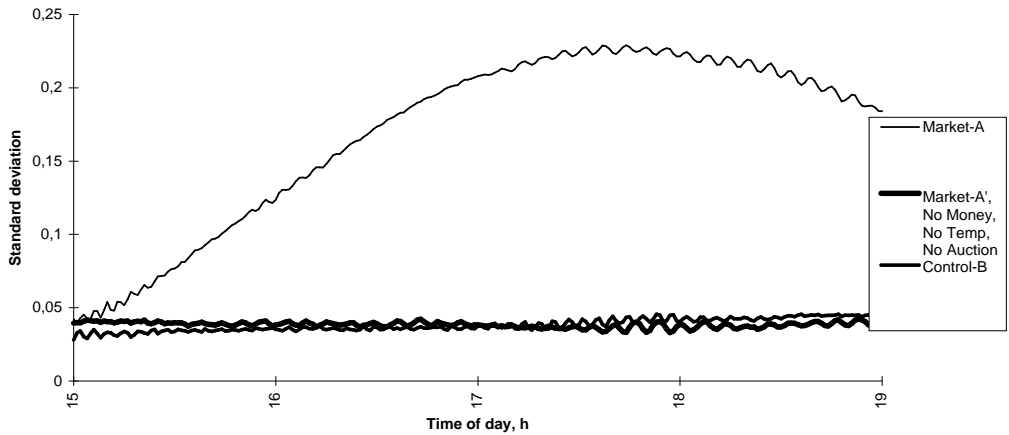

Figure 12: *Standard deviation with an integrating controller that utilizes global data compared to the Huberman-Clearwater* Market-A *scheme, and to the* Market-A′ *scheme with the auction removed.*

## 9. Market-B: A Market with Local Data Only

Thus, the computational market Market-A is outperformed by the global control scheme Control-B. Therefore, an interesting follow-up question is whether a simple and well performing computational market approach can be devised that does *not* depend on having available global information, in contrast to both these schemes. In this section we show that this is indeed the case, such that a decentralized market performs comparably to fully centralized conventional control. The approach derived in this section relies on more general theory for the relations between optimization problems and markets–(Ygge, 1998, Chapter 3) and (Bikhchandani & Mamer, 1997) – applicable to this and many other applications.

---

13. This holds under the assumption that the characteristic time scale of the variations of the average temperature is much larger than that of the fluctuation of the temperature. This is a very reasonable assumption in the present case.

We use a market-oriented programming approach based on an equilibrium market. The basic protocol for this is (cf. Figure 1):

1. Each agent submits a *net demand function $z_{io}(p)$* to the auctioneer, describing what change in allocation is desired by agent $o$ at price $p$.

2. The auctioneer computes an equilibrium, by calculating a price $p_i$ such that $\sum_o z_{io}(p_i) = 0$ (or, strictly speaking, $|\sum_o z_{io}(p_i)| \leq \epsilon$, where $\epsilon$ is a small (positive) numerical tolerance).

3. Each agent receives its demanded resource $z_{io}(p_i)$ as calculated at the obtained market equilibrium price.

The computation of $z_{io}(p)$ and the process of computing the equilibrium are described further below. Note that this is a truly distributed approach, as it does not rely on any global data, or communication of temperatures between the agents before the market communication starts.

## 9.1 The Relation Between Markets and Conventional Controllers utilizing Global Data

The performance measure for the system is given by Eq. (1). The best system is therefore one that minimizes this equation. Hence, the most straightforward move one can think of to come to a market model, is to take *this* measure as representing the utility function for the overall system. So, the utility functions for the individual agents are ideally related to $[(T_{io} - T_o^{setp}) - (\langle T_i \rangle - \langle T^{setp} \rangle)]^2$. [14] However, this is still a formulation containing global information.

Thus, we want to obtain a *purely local* reformulation, by getting rid of the terms with $\langle T \rangle$ containing the global information. We might replace them, though, by terms relating to the changes in the local resource. In doing so, we take inspiration from the standard controller equations Eqs. (11) and (18), indicating that we get good results (for unconstrained resources) with the update equation $F_{io} = F_{i-1,o} + \phi_{io}$, when $\phi_{io}$ has the form $\phi_{io} = \beta(T_{io} - T_o^{setp})$. The intended interpretation of $\phi_{io}$ is to represent the output that the local controller would have delivered if it acted independently with unconstrained resources. In the market setting each agent updates its control signal $F_{io}$ through $F_{io} = F_{i-1,o} + \Delta F_{io}$, where $\Delta F_{io}$ is determined by the outcome of the market. Since the resource is only redistributed among the agents, we have that $\sum_{o=1}^{N} \Delta F_{io} = 0$. Accordingly, as a step in the design of a MARKET-B scheme, based on local data only, we employ the following definition.

**Definition 9.1** *Let the utility function for the individual office agents be defined by*

$$u(\Delta F_{io}, m) = -\alpha_o^2 (\Delta F_{io} - \phi_{io})^2 + m, \tag{19}$$

*where $\alpha_o$ is a strength parameter for each office representing its physical properties such as $R_o$ and $C_o$. (The proper choice of $\alpha_o$ is discussed in Theorem 9.3 later on.)*
*Furthermore, let all agents be price-takers.*

---

14. Since utilities are expressions of preference orderings, they are invariant under monotonic transformations.

Due to the simple quadratic form of the utility function in Eq. (19), the net demand, $z_{io}(p)$, is easily computed for price-taking agents by analytical means. Note that the money $m$ here are fundamentally different from the money of MARKET-A, c.f. Section 6. Here the money represents a commodity and the agent needs to decide how much money to trade for the commodity cold air. As described above, the money in Section 6 is just a control variable without an intuitive interpretation.

**Theorem 9.1** *With the utility functions and behavior given by Definition 9.1, there exists a unique equilibrium price $p_i | \sum_o z_{io}(p_i) = 0$, and the allocation obtained by the equilibrium market is Pareto optimal.*

**Proof.** See (Mas-Colell, Whinston, & Green, 1995; Varian, 1992; Takayama, 1985). □

In this setting, there is a wide variety of different algorithms that are guaranteed to converge to the clearing price $p_i | \sum_o z_{io}(p_i) = 0$, in the second step of our equilibrium market protocol given above. Examples of such algorithms are price tâtonnement, WALRAS (Cheng & Wellman, 1998), and various Newton-Raphson type methods. A detailed discussion of communication and computation efficiency aspects of such algorithms is given by Ygge (1998, Chapter 4).

**Theorem 9.2** *Any Pareto optimal outcome in a market in which the agents hold the utility function of Eq. (19)* **is equivalent to** *an integrating controller that exploits global information as described by the following resource update equation :*

$$\Delta F_{io} = \phi_{io} - \frac{1}{\alpha_o^2 \cdot \langle 1/\alpha^2 \rangle} \langle \phi_i \rangle, \tag{20}$$

*given that no agent is at its boundary. Here, $\langle 1/\alpha^2 \rangle$ and $\langle \phi_i \rangle$ are the average $1/\alpha_o^2$, and $\phi_{i0}$ respectively.*

**Proof.** At a Pareto-optimal allocation where no agent is at its bounds, all $\partial u(\Delta F_{io})/\partial \Delta F_{io}$ are equal. (Assume $\partial u(\Delta F_{ip})/\partial \Delta F_{ip} < \partial u(\Delta F_{iq})/\partial \Delta F_{iq}$, then at any price $\partial u(\Delta F_{ip})/\partial \Delta F_{ip} < p < \partial u(\Delta F_{iq})/\partial \Delta F_{iq}$, there exists a (sufficiently small) amount of $\Delta F_i$, say $\epsilon$, such that $u_p(\Delta F_{ip} - \epsilon, m_{ip} + p\epsilon) > u_p(\Delta F_{ip}, m_{ip})$ and $u_q(\Delta F_{iq} + \epsilon, m_{iq} - p\epsilon) > u_q(\Delta F_{iq}, m_{iq})$. Correspondingly for $\partial u(\Delta F_{ip})/\partial \Delta F_{ip} > \partial u(\Delta F_{iq})/\partial \Delta F_{iq}$. Hence, $\partial u(\Delta F_{ip})/\partial \Delta F_{ip} = \partial u(\Delta F_{iq})/\partial \Delta F_{iq}$.

Thus, it will hold for every office that $\Delta F_{io} - \phi_{io} = \frac{\alpha_N^2}{\alpha_o^2}(\Delta F_{iN} - \phi_{iN})$. Summing the equations yields $\sum_{o=1}^{N-1} \Delta F_{io} - \sum_{o=1}^{N-1} \phi_{io} = \alpha_N^2(\Delta F_{iN} - \phi_{iN}) \sum_{o=1}^{N-1} \frac{1}{\alpha_o^2}$. Adding $\Delta F_{iN} - \phi_{iN}$ to both sides and dividing both sides by $N$ together with the resource constraint ($\sum_{o=1}^{N} \Delta F_{io} = 0$) yields $\frac{-\sum_{o=1}^{N} \phi_{io}}{N} = \alpha_N^2(\Delta F_{iN} - \phi_{iN})\frac{\sum_{o=1}^{N} \frac{1}{\alpha_o^2}}{N}$ which is equivalent to $\Delta F_{iN} = \phi_{iN} - \frac{1}{\alpha_N^2 \cdot \langle 1/\alpha^2 \rangle}\langle \phi_i \rangle$. For reasons of symmetry, this equation holds for all offices. □

**Corollary 9.1** *For the special case where all $\alpha_o$ are equal, and $\phi_{io} = \beta(T_{io} - T_o^{setp})$, Eq. (20) becomes exactly the CONTROL-B scheme captured by Eq. (18).*

Consequently, it also follows that simulation results for an equilibrium Market-B scheme, based on the utility function of Eq. (19) with the provisions given in the above corollary, *are identical* to those of the centralized Control-B scheme, cf. Figure 12. This fully decentralized market scheme thus performs better than the original Huberman-Clearwater market protocol.

In sum, we see that (under suitable conditions which are fulfilled here) *local data plus market communication is equivalent to conventional central control utilizing global data.* Even though our proof was based on the assumption that the agents are never at their boundary values, it will be a close approximation in many practical applications. It should also be noted that managing the boundaries is not required for a successful implementation. As seen above, omitting the management of boundaries in the current application leads to Control-B, which was shown to have a very high performance.

## 9.2 Finding an Optimal Utility Function

In this section we show how an optimal utility function is constructed in the *constrained* case, from an optimal controller for the *unconstrained* case.[15] Earlier we noted that it is debatable whether or not the measure in Eq. (1) is a good one. One argument against it is that it allows for pathological solutions, as it does not take the average indoor temperature into consideration. If we for example increase the indoor temperature in every office by $10°C$, all of them could be unbearably hot, while the measure might be minimized. A reasonable thing to do here, to prevent these kind of solutions, is to treat the average temperature as given (the result of using all available resources). If we do so, the following theorem shows how an optimal utility function is constructed in the constrained case, from an optimal controller for the unconstrained case.

**Theorem 9.3** *If $T_{io}$ is a linear function of $\Delta F_{io}$,[16] and if $\phi_{io}$ minimizes Eq. (1) in the unconstrained case, then in a market where the utility functions are described by Eq. (19), with $\alpha_o = \frac{\partial T_{io}}{\partial \Delta F_{io}}$, any associated Pareto optimal allocation minimizes Eq. (1) in the constrained case, if the resource can be independently allocated among the agents (i.e., when the only constraint on $\Delta F_{io}$ is the upper and lower bound, and $\sum_{o=1}^{N} \Delta F_{io} = 0$), and the average temperature is considered as given.*

**Proof.** Minimizing Eq. (1) boils down to minimizing $f(T_{i1}, T_{i2}, \dots, T_{in}) = \sum_{o=1}^{N} [(T_{io} - T_o^{setp}) - (\langle T_i \rangle - \langle T^{setp} \rangle)]^2$, since this is just a monotonic transformation. Due to the fact that $\phi_{io}$ minimizes Eq. (1), it holds that $\frac{\partial f}{\partial \Delta F_{io}} = 0$ for $\Delta F_{io} = \phi_{io}$. Since the average temperature is considered as given, we have that $\frac{\partial f}{\partial \Delta F_{io}} = 2[(T_{io}(\phi_{io}) - T_o^{setp}) - (\langle T_i \rangle - \langle T^{setp} \rangle)] \cdot \frac{\partial T_{io}}{\partial \phi_{io}}$ for $\Delta F_{io} = \phi_{io}$. This gives that $T_{io}(\phi_{io}) = T_o^{setp} + (\langle T_i \rangle - \langle T^{setp} \rangle)$. Thus, $f$ can be rewritten as $\sum_{o=1}^{N} [T_{io}(\Delta F_{io}) - T_{io}(\phi_{io})]^2$. Since $T_{io}$ is a linear function of $\Delta F_{io}$, we have that $T_{io}(\Delta F_{io}) = T_{io}(\phi_{io}) + \frac{\partial T_{io}}{\partial \Delta F_{io}} \cdot (\Delta F_{io} - \phi_{io})$, and hence $f$ becomes $\sum_{o=1}^{N} \left( \frac{\partial T_{io}}{\partial \Delta F_{io}} \right)^2 (\Delta F_{io} - \phi_{io})^2$. As the reallocation of $m$ does not effect the total (summed) utility, minimizing $f$ is equivalent to minimizing $-\sum_{o=1}^{N} u_o$, with $\alpha_o = \frac{\partial T_{io}}{\partial \Delta F_{io}}$. Furthermore, we have that any Pareto optimal allocation in a market where the utility functions are described by Eq. (19) maximizes $\sum_{o=1}^{N} u_o$. (Suppose that there is an allocation $\Delta F_{io}^a$ that does not maximize $\sum_{o=1}^{N} u_o$, but that $\Delta F_{io}^b$ does. Then reallocating $\Delta F_{io}$ to $\Delta F_{io}^b$ and letting $m_{io}^b = m_{io}^a + u_o(\Delta F_{io}^a, m_{io}^a) - u_o(\Delta F_{io}^b, m_{io}^b) + \frac{\sum_{o=1}^{N} u_o(\Delta F_{io}^b, m_{io}^b) - \sum_{o=1}^{N} u_o(\Delta F_{io}^a, m_{io}^a)}{N}$ is always a Pareto improvement. Hence, a non-maximized sum implies a non-Pareto optimal allocation, and a Pareto optimal allocation implies a maximized sum.) Then (Ygge, 1998, Theorem 3.1, p. 44) implies that any Pareto-optimal allocation in a market where the utility functions are described by Eq. (19) with $\alpha_o = \frac{\partial T_{io}}{\partial \Delta F_{io}}$, minimizes $f$. $\square$

---

15. The resource is constrained when the action of any agent is limited by $P_{io}$, see Eq. (12) and Figure 7. It is unconstrained when such a limitation is not present, cf. Figure 6.

16. From the thermal model discussed in Section 4 (especially Eqs. (5) and (12)) we note that this is indeed the case for a reasonably wide range.

Thus, we have constructed a fully distributed market design that yields an optimal outcome. Particularly note that Theorem 9.3 was not based on the assumption that an integral controller was used, rather it said that if $\phi_{io}$ (the desired resource by the controller) is optimal, then the proposed utility function generates a globally optimal outcome.

## 9.3 Discussion

Previously, we saw that the independent controller CONTROL-B that incorporates global data, viz. the average temperatures, performs very well. In the present section we positively answered the question if one can construct a market, MARKET-B, that is based on local data only and that performs as well.

For this result we have employed a market approach based on general equilibrium theory. This is of course not the only available mechanism for resource allocation in multi-agent systems. It seems interesting to try out other mechanisms, like the contract net protocol (Davis & Smith, 1983), and see if they perform better. However, Theorem 9.3 tells us that, if we treat the problem of building control as being separable in terms of agents (as also done by Huberman and Clearwater), there is no better scheme.[17] For example, if we assigned all the resources to an auctioneer, that on its turn would iteratively assign the total resource in small portions to bidders bidding with their true marginal utility, we would end up with something close to the competitive equilibrium, but we cannot do better than MARKET-B. Furthermore, this would be a computationally extremely inefficient way to arrive at equilibrium compared to other available methods (Ygge, 1998, Chapter 4). That is, we can use different mechanisms for achieving the competitive equilibrium, but we can never hope to find a mechanism that would do better than the MARKET-B scheme.

---

17. We note that, as mentioned earlier, the problem is actually not fully separable in terms of agents and therefore better solutions may exist if this is taken into account. Another observation is that in this article we have investigated only the case of using the currently available resource as the interesting commodity, in accordance with the work of Huberman and Clearwater, and we found an optimal mechanism for that. We note however that extending the negotiations to future resources as well could potentially increase performance. But this is a different problem setting with different demands on available local and/or global information items, such as predictions. We have given a solution to this problem in recent work, see Ygge et al. (1999).

## 10. Conclusions

We believe that both the approach and the results, as presented in this article, pose an interesting challenge to the software agent community. Multi-agent systems offer a new way of looking at information systems development, with a potentially large future impact. However, new approaches must prove their value in comparison and competition with existing, more established ones. The agent paradigm is no exception.

We have therefore deliberately played the role of the devil's advocate in this article. In our view, a key question not yet satisfactorily answered by the software agent community is: in what respect and to what extent are multi-agent solutions better than their more conventional alternatives? This article has shown that arguing in favor of the multi-agent systems approach does require careful analysis beyond the disciplinary boundaries of computer science. Empirical and comparative studies in the multi-agent literature of the kind carried out in the present article are all too rare as yet. But as we have shown in technical detail, established paradigms such as conventional central control cannot be that easily dismissed. A similar argument holds, by the way, regarding mathematical optimization techniques in distributed resource allocation problems, cf. our previous discussions (Ygge & Akkermans, 1996; Ygge, 1998). Many of them are, in a distributed guise, better than many newly proposed market protocols.

Abstract considerations alone, concerning the general nature of agenthood, autonomy, rationality, or cooperation, are not sufficient to prove the value of the agent paradigm. Such theoretical reflections are worthwhile, but do not diminish the need for thorough analysis of (agent and market) failures *and* successes in real-life applications. Therefore, we have taken a different approach, aimed at obtaining experimental data points on the basis of which convincing software agent claims can be established.

The data point considered in this article is climate control within a large building consisting of many offices. Given a measure, Eq. (1), a local controller approach, Eq. (11), current temperatures, and the setpoint temperature, we have investigated the problem on how to properly distribute the resource among the offices. This is a rather prototypical application relating to the general problem of optimal resource allocation in a highly distributed environment. This class of problems has already received much attention in the multi-agent area. Reportedly, this type of application is very suitable for market-oriented programming (Huberman & Clearwater, 1995). On the other hand, we have devised some better conventional control engineering solutions, as well as alternative and better market designs.

The main conclusions of our investigation are:

- The market approach by Huberman & Clearwater (1994, 1995) indeed outperforms a standard control solution based on local, independent controllers. So, the market-based multi-agent approach indeed yields a working solution to this type of problem. Our analysis has shown that the success of this market approach depends on the agents communicating their local information to all other agents before the auction starts.

- However, if conventional central control schemes are allowed to exploit the same global information, they perform even better.

- We have proposed an alternative market design based on general equilibrium theory that uses local data only. It performs better than the Huberman-Clearwater market and as well as a centralized control scheme having access to global information.

- Our general conclusion can be formulated as a quasi-equation: *"local information + market communication = global control"*. This holds under suitable conditions (existence of market equilibrium, price-taking agents) and its validity has been specifically shown here for the case of building climate control (compare particularly the CONTROL-B and the MARKET-B schemes we developed). However, as we intend to show in forthcoming work, this is a much more generally valid result.

- The important difference is that in computational markets this global information is an *emergent property* rather than a presupposed concept, as it is in central control.

In our analysis we have focused on the market approach. It is tempting to ask whether things are different when a non-market multi-agent approach is followed, say, using the contract net (Davis & Smith, 1983). As we argued, the answer in our opinion is a straightforward *no*. The goal in the considered class of problems is to find the optimal distributed solution. Alternative agent approaches, market as well as non-market ones, only change the multi-agent dynamics on the way to this goal. This might be done in a better or poorer way, but it is not possible to change this goal itself. The goal state in any multi-agent approach is, however formulated, equivalent to market equilibrium, the yardstick for having achieved it is given by some quantitative performance measure as we discussed, and both are stable across different agent approaches.

Thus, one of the main conclusions of this article is that one must be very careful when promoting multi-agent approaches to resource allocation problems for which traditional approaches are available. In particular, one should be cautious using arguments related to distribution of information (as was done by Huberman and Clearwater), or other computational aspects. Still, we argue that there *are* conceptual advantages of market-based approaches to this type of problems, as well as other advantages such as evaluation of local performance (cf. Section 2). These advantages do not show off very clearly in the present application setting – for example, it was assumed that all offices have the same thermal characteristics and office agents on the market are not added and deleted continuously on the fly – but they are clearly visible in other more general settings.

A final note is that in this article we have devised an *optimal market design given the problem formulation* rather than devised *an optimal approach to the general problem of building control*. This is in line with the objective of this paper to give a comparative study of different possible approaches. That is, we have focused on a published and well-known problem formulation in order to focus on the subject of markets for resource allocation alone. As stated earlier in the article, several aspects can be improved, for example:

- The local controllers. I-controllers are seldom used, rather PI- or PID-controllers are preferred. Furthermore, for these kind of computerized settings, modern digital controllers, based on e.g. pole-placement methods are preferable (Åström & Wittenmark, 1990).

- The measure. The average value is at least as important as the standard deviation. (We have as a consequence excluded pathological solutions from our analysis by taking the average temperature as given in Theorem 9.3).

- The incentives for office agents to reveal their true preferences in order to avoid speculation. Reasonably, personnel in the offices should have the opportunity to make trade-offs between comfort and economical value. (It has been shown that generally speaking it is very difficult to benefit from speculation if the number of agents on the market is sufficiently large, and/or if uncertainty about the other agents' behavior exists (Sandholm & Ygge, 1997).)

- Taking several time periods into account. As each office serves as a storage for heat, agents can for example gain by using relatively much resource during certain hours when the total resource need is small.

That is, a more realistic approach to the building control problem is to use minimized cost as the measure, give the users real incentives to reveal their energy preferences (by letting them pay their actual energy costs so that attempted speculation will generally cost money), and take future time periods into account. Elsewhere we have solved the problem of dealing with simultaneous optimization over different time periods, by means of a multi-commodity market (Ygge et al., 1999).

In such a more realistic and large-scale setting, the market-based approach is attractive compared to its alternatives. The abstractions used (prices and demands) are natural and easily understood by everyone, and they are uniform over all types of agents (even ones that do not use the resource for controlling a temperature).

A major qualitative conclusion of this article is that "local data plus market communication yields global control". This conclusion is based on the discussed application setting of building control, as an instance of a distributed resource allocation problem. It suggests that generally for this type of problem both central control engineering and multi-agent market solutions can be devised that give comparable optimal control quality. A next step is to prove this more generally in a rigorous mathematical fashion, delineating the preconditions in more detail. Such a proof can be based upon the (continuous and matrix-algebraic) dynamic systems theory available from control engineering, and especially upon results concerning what is called optimal control. We believe that this indeed can be done in a formal way, and *as a conjecture we state the following general result.* Multi-agent equilibrium markets yield optimal decentralized solutions to distributed resource allocation problems that are of the same quality, in terms of a given overall systems performance index, as solutions given by the optimal central (multi-input, multi-output) systems controller that has access to all relevant local data. These local data involve the total system state and control vectors (in building control these are the vectors formed from the difference between the actual and setpoint temperatures for each office, and the cooling power for each office, respectively). Preconditions for this to be true are: (i) agents act competitively; (ii) the equilibrium exists; (iii) the systems performance index can be written as a linear combination of local contributions (implying diagonality of certain matrices; if not, agents are not independent). These local contributions are directly related to the agents' utility functions. The agent approach will be more readily generalizable to large-scale systems and non-linear control solutions (linearity is the main case where control engineering can get analytical mathematical expressions). This is indeed a strong and general statement about the relationship (and even outcome equivalence) between decentralized markets and central control.

## Acknowledgments

The present work was mainly carried out at the University of Karlskrona/Ronneby. We thank Rune Gustavsson and Hans Ottosson for all their support. A special acknowledgment goes to Olle Lindeberg whose very detailed comments on the draft papers led to significant improvements in the simulations and whose ideas also helped us in the design of the market in Section 9. We benefit significantly from several discussions with Michael Wellman. We also thank Bengt Johannesson, who, as a part of his master's thesis, went through all details of this article and independently recreated and verified all the simulations. We thank Arne Andersson, Eric Astor, and the SoC team for useful comments and discussions of draft material. This work was partially sponsored by NUTEK and EnerSearch AB.

An earlier version of this paper was presented at the MAAMAW'97 workshop (Ygge & Akkermans, 1997).

## Appendix A. Original Formulation of the Huberman-Clearwater Market

All formulae in this section were directly taken from the papers by Huberman & Clearwater (1994, 1995).

**Trade volumes**  First, the decision for an agent to buy or sell is based on

$$t_{io} = \frac{T_o^{setp}}{T_{io}} \cdot \frac{\langle T_i \rangle}{\langle T^{setp} \rangle} \quad \left\{ \begin{array}{ll} t_{io} > 1, & seller \\ t_{io} < 1, & buyer \end{array} \right. . \tag{21}$$

Then, the total trade volume, $V$, is calculated from

$$V_i = \sum_{o=1}^{N} |1 - t_{io}|, \tag{22}$$

where N is the number of offices.

Every agent calculates its request volume, $v$, according to

$$v_{io} = \alpha \frac{|1 - t_{io}|}{V_i}. \tag{23}$$

When an agent buys or sells its $v$ the actual movement of a valve, called $VAV$, is computed from

$$\Delta VAV_{io} = f(flow_{io}, v_{io}, VAV_{io}), \tag{24}$$

after which the actual $VAV$ position for each interval is updated according to

$$VAV_{i+1,o} = VAV_{io} + \Delta VAV_{io}. \tag{25}$$

The function $f$ in Eq. (24) stems from the physics of the office control, but it is not easy to derive in the general case and it is neither explicitly given in the original papers (Clearwater & Huberman, 1994; Huberman & Clearwater, 1995). A reasonable and simple choice is to assume a linear relationship (the cited papers suggest that they do so in practice as well), replacing Eq. (24) and Eq. (25) by

$$F_{i+1,o} = F_{io} \pm v_{io}, \tag{26}$$

where plus or minus depend on whether it refers to an accepted buy or sell bid. $P_{cio}$ is obtained from Eq. (12). We have employed Eq. (26) in our simulations. The nature of this assumption is not crucial to our central line of reasoning.

**Bids**  The bids are based on a *marginal utility*[18] of the form described by[19]

$$U(t_{io}/T_o^{setp}, m_{io}) = [U(0, m_{io})]^{(1-t_{io}/T_o^{setp})} = [U(0, m_{io})]^{\left(1-\frac{\langle T_i \rangle}{T_{io}\langle T^{setp} \rangle}\right)}, \tag{27}$$

---

18. This is called *utility* in the work of Huberman and Clearwater. We prefer to use the term marginal utility instead, because it is directly related to price. This terminology conforms better to that of microeconomic theory.

19. This notion of a utility function used by Clearwater and Huberman is very much based on the work of Steiglitz & Honig (1992)

with

$$U(0, m_{io}) = u_3 - (u_3 - u_1)e^{-\gamma m_{io}}, \tag{28}$$

and

$$\gamma = ln\left[\frac{u_3 - u_1}{u_3 - u_2}\right], \tag{29}$$

where $u_1 = 20$, $u_2 = 200$, and $u_3 = 2000$, and $m$ is the amount of money that an agent has, given by

$$m_{io} = 100(2 - VAV_{io}). \tag{30}$$

With the above relation between $F_{io}$ and $VAV$, we rewrite Eq. (30) as

$$m_{io} = 100(2 - \frac{F_o^{max} - F_{io}}{F_o^{max}}). \tag{31}$$

Observing Eqs. (28) and (29), we note that these equations can be simplified to

$$\begin{aligned} U(0, m_{io}) = \ & u_3 - (u_3 - u_1)e^{-\gamma m_{io}} = u_3 - (u_3 - u_1)(e^\gamma)^{-m_{io}} = \\ & u_3 - (u_3 - u_1)\left(\frac{u_3 - u_1}{u_3 - u_2}\right)^{-m_{io}}. \end{aligned} \tag{32}$$

The bids are calculated from multiplying the marginal utility with the previous price, $price$, according to

$$B_{i,o} = U_{io}(t_{io}/T_o^{setp}, m_{io}) \cdot price_i. \tag{33}$$

This is the equation given by Huberman and Clearwater. Straightforward application of Eq. (33) turns out to produce major problems in simulations, however. Equation (27) shows that $U(t_{io}/T_o^{setp}, m_{io})$ is minimized for minimized $T_{io}$ and maximized $\frac{\langle T_i \rangle}{\langle T^{setp} \rangle}$. As we can expect $T_{io}$ to be well above $10°C$ and $\frac{\langle T_i \rangle}{\langle T^{setp} \rangle}$ to be well below 2, and $U(0, m) \approx 2000$, we can be sure that $U(t_{io}/T_o^{setp}, m_{io})$ will be well above $2000^{1-\frac{2}{10}} \approx 437$. Thus, the bidding price will never be below 437. Then, Eq. (33) tells us that the market price will be at least $price_0 \cdot 437^i$. This leads to numerical overflow after a few iterations. We note however that, since all agents multiply their bids with the previous price, this has no effect on the reallocation itself: it affects only the price level. Therefore, we omit multiplying by the previous price in our simulations, so that the agents' bid prices in Eq. (33) equal their marginal utilities of Eq. (27). Huberman and Clearwater state that they burn all money after each auction round in order to avoid inflation and overflow, indicating that they may adopt a similar procedure. In any case, this procedure leads to exactly the same allocations but avoids numerical overflow.